\documentclass[jkps,twoside,twocolumn]{revtex4}
\usepackage{amssymb,amsfonts,graphicx}

\begin{document}
\title[]{Effects of the globalization in the Korean financial markets}
\author{Woo-Sung \surname{Jung}}
\email{wsjung@kaist.ac.kr}
\thanks{Fax: +82-42-869-2510}
\author{Okyu \surname{Kwon}}
\author{Jae-Suk \surname{Yang}}
\author{Hie-Tae \surname{Moon}}
\affiliation{Department of Physics, Korea Advanced Institute of
Science and Technology, Daejeon 305-701, Republic of Korea}
\received{September 13, 2005}

\begin{abstract}
We study the effect of globalization on the Korean market, one of
the emerging markets. Some characteristics of the Korean market are
different from those of the mature market according to the latest
market data, and this is due to the influence of foreign markets or
investors. We concentrate on the market network structures over the
past two decades with knowledge of the history of the market, and
determine the globalization effect and market integration as a
function of time.
\end{abstract}


\maketitle


\section{Introduction}
`The world to Seoul, Seoul to the world.' This was the slogan of the
1988 Seoul Olympics Games, and is also the slogan of the Korean
stock market. The globalization means that foreign traders have an
influence on the Korean market and its synchronization with world
markets.

Interdisciplinary study has received much attention, with
considerable interest in applying physics to economics and finances
\cite{stanley,arthur,bouchaud,aste,kaizoji}. Since a financial
market is a complex system, many researchers have developed network
theory to analyze such systems. The concept of an asset tree
constructed by a minimum spanning tree is useful in investigating
market properties \cite{mantegna,bonanno,johnson}. The minimum
spanning tree (MST) is derived for a unique sub-network from a fully
connected network of the correlation matrix. The MST of $N$ nodes
has $N-1$ links; each node represents a company or a stock and edges
with the most important correlations are selected. Then clusters of
companies can be identified. The clusters, a subset of the asset
tree, can be extended to portfolio optimization in practice. The
companies of the US stock market are clearly clustered into business
sectors or industry categories \cite{onnela}.

Nowadays, many emerging markets experience the globalization that is
making rapid progress, and the influence of developed markets is
becoming stronger. Most markets synchronize with the US market and
globalization is leading to characteristic changes in emerging
markets \cite{climent}.

Several results have been reported on the necessity to find a model appropriate
to emerging markets, because the models for mature
markets cannot be applied to emerging markets universally
\cite{india}. The Korean market is representative of emerging
markets and is subject to synchronization with external markets
\cite{jkps1,jkps2,jkps3,jkps4,jkps5}. Clustering in the
Korean market differs from that in the US market and is due to
foreign factors \cite{wsjung}.

In this paper, we explore characteristics of the Korean stock
market. We construct the minimum spanning tree (MST) shifting a time
window of approximately two decades and analyze the time-dependent
properties of the clusters in the MST that the market conditions are
not stationary. Then we investigate the market with knowledge of the
history of the Korean market.

\section{Dynamics of the market}
The Korea Stock Exchange (KSE) opened in 1956. At that time, only 12
companies were listed on the market. As the Korean economy has
developed, the stock market has undergone many changes under the
influence of factors inside and outside the market.

\begin{figure}[h]
\includegraphics[angle=0,width=7cm]{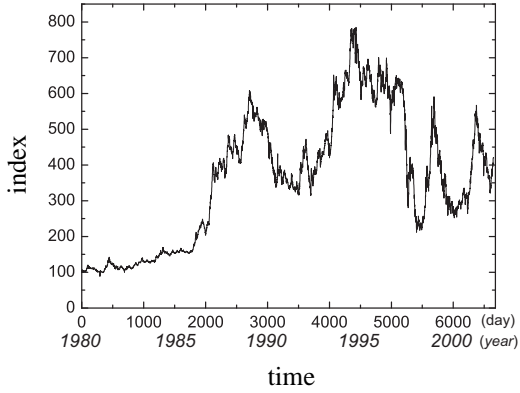}
\caption[0]{Index of
228 selected companies in the Korean stock market from 1980 to
2003. } \label{index}
\end{figure}

We deal with the daily closure stock prices for companies listed on
the KSE from 4 January 1980 to 30 May 2003. The stock had a total of
6648 price quotes over the period. We select 228 companies that
remained in the market over this period of 23 years. Fig.
\ref{index} shows the index for those companies. The representative
KSE index, KOSPI, is an index of the value-weighted average of
current stock prices. The index of Fig. \ref{index} is a
price-equally-weighted index, similar to use for the Dow Jones
industrial average (DJIA). Many previous studies on the stock market
assumed a certain number of trading days to constitute a year.
However, it is not easy to apply such an assumption to our data set,
because the Korean market opening time changed in 2000. Before 20th
May 2000, the market opened every day except Sunday, and from Monday
to Friday after 21th May 2000. Most of data set falls into the
former period, so we assume 300 trading days for one year. The
x-axis values in Fig. \ref{index} were calculated under this
assumption.

We use the logarithmic return of stock $i$, which can be written as:

\begin{equation}
S_i(t) = \ln Y_i(t+\Delta t)-\ln Y_i(t),
\end{equation}

\noindent where $Y_i(t)$ is the price of stock $i$. The cross-correlation
coefficients between stock $i$ and $j$ are defined as:

\begin{equation}
\lambda_{ij} =\frac{<S_{i}S_{j}>-<S_{i}><S_{j}>}
{\sqrt{(<S_{i}^{2}>-<S_{i}>^{2})(<S_{j}^{2}>-<S_{j}>^{2}) }}
\end{equation}

\noindent and form a correlation matrix $\Lambda$.

\begin{figure}[h]
\includegraphics[angle=0,width=7cm]{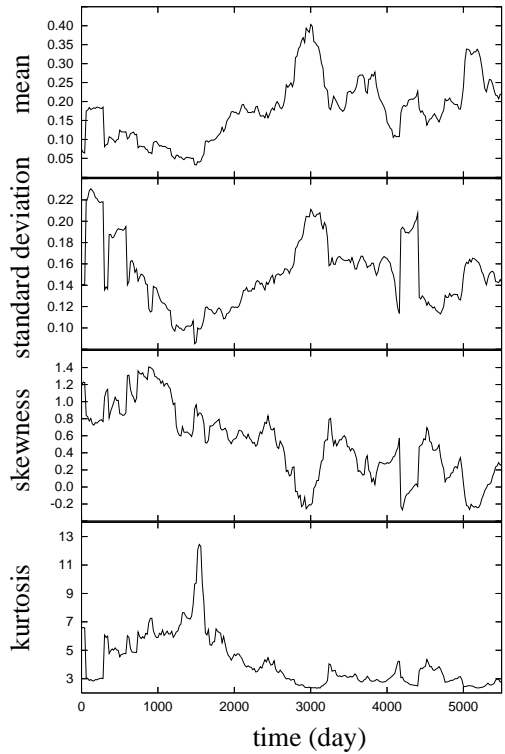}
\caption[0]{The mean, standard deviation, skewness, and kurtosis of
the correlation coefficient in the Korean market as functions of
time.} \label{correlation}
\end{figure}

\begin{table}[t!]
\caption{Industry categories of the Korea Stock Exchange in our data
set} \label{category}
\begin{tabular}{|p{1.5cm}|p{4cm}|p{1.5cm}|}
\hline
Category number&Industry category&No. of companies\\
\hline
1&Fishery \& Mining&1\\
2&Food \& beverages&24\\
3&Tobacco&0\\
4&Textile&14\\
5&Apparel&3\\
6&Paper \& wood&10\\
7&Oil&0\\
8&Chemicals \& medical supplies&40\\
9&Rubber&6\\
10&Non-metallic minerals&12\\
11&Iron \& metals&10\\
12&Manufacturing \& machinery&13\\
13&Electrical \& electronic equipment&8\\
14&Medical \& precision machines&1\\
15&Transport equipment&12\\
16&Furniture&0\\
17&Electricity \& gas&1\\
18&Construction&21\\
19&Distribution&17\\
20&Transport \& storage&10\\
21&Banks&8\\
22&Insurance&11\\
23&Finance&4\\
24&Services&1\\
25&Movies&1\\
\hline
\end{tabular}
\end{table}

The top panel of Fig. \ref{correlation} shows the mean correlation
coefficient calculated with only non-diagonal elements of $\Lambda$.
The second shows the standard deviation, the third, the skewness and
the last, the kurtosis. It has been reported that when the market
crashes, the correlation coefficient is higher \cite{drozdz}. In the
US market, the effect of Black Monday (19 October 1987) was clearly
visible for these four coefficients, with correlations among them
also apparent \cite{onnela}. However, crash effects on the Korean
market (the late 1980s bubble crash and the 1997 Asian financial
crisis) are visible, but not clear in comparison with the US market,
and the Korean market coefficients do not have clear correlations.

We investigate more properties of the market through the MST that is
a simple graph with $N$ nodes and $N-1$ links. The most important
connection is linked when it is constructed. It is known that the US
market network is centralized to a few nodes \cite{kim}. The hub of
the US market is approximately General Electric (GE), and it is
possible to make clusters (subsets of the MST) of the US market with
industry categories or business sectors \cite{onnela}. However, the
Korean market has no comparable hub for the whole market, and the
clusters are constructed with the MSCI index \cite{wsjung}. We
regard this result as the effect of globalization and market
integration. Thus, we obtained the MSTs from 1980 to 2003 with time
windows of width $T$ corresponding to daily data for $T$=900 days
and $\delta T$=20 days. During this period there is no comparable
hub, but we can form clusters with industry categories for some
periods. Then we define the parameter \emph{grouping coefficient}.
The grouping coefficient of a specified industry category $C$ is
defined as:

\begin{equation}
g_{C}=\frac{n^{C}(i_{\in C})}{n(i_{\in C})},
\end{equation}

\noindent where $i_{\in C}$ represents the nodes in category $C$,
$n(i)$ is the number of links that are connected to node $i$ and
$n^{C}$ is the number of links from the node included in category
$C$.

\begin{figure}[b]
\includegraphics[angle=0,width=8cm]{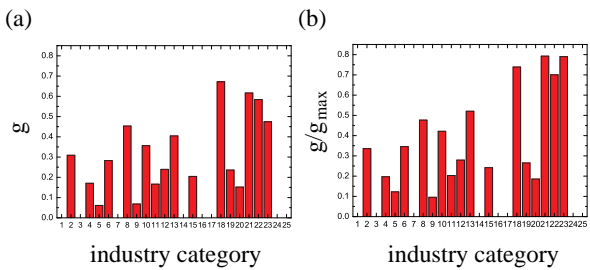}
\caption[0]{Plot of grouping coefficients: (a) shows the value of
$g$ and (b) shows the ratio of the coefficient to the maximum value
of $g$.} \label{group}
\end{figure}

Table \ref{category} shows 25 industry categories of the Korean
stock market. In fact, there are rather more categories than 25.
However, the standard for grouping is excessively detailed, and
combinations of these categories are mostly used. The categories in
Table \ref{category} are reconstructed from Hankyoreh, a popular
newspaper in Korea. Fig. \ref{group}a shows the \emph{grouping
coefficient} for each category over the whole period. We observe
that categories 8, 18, 21, 22 and 23 form a well-defined cluster. We
focus on the maximum grouping coefficient for each industry
category. For example, there are only four companies in the finance
category (23) and the maximum value of the coefficient is only 0.6
(=3/5) because of the characteristics of the MST. We take the
maximum value when the nodes are linked linearly. Fig. \ref{group}b
shows the ratio of the grouping coefficient to the maximum value for
each category. Categories 18, 21, 22 and 23 are almost complete
clusters in this plot.

\begin{figure}[h]
\includegraphics[angle=0,width=7cm]{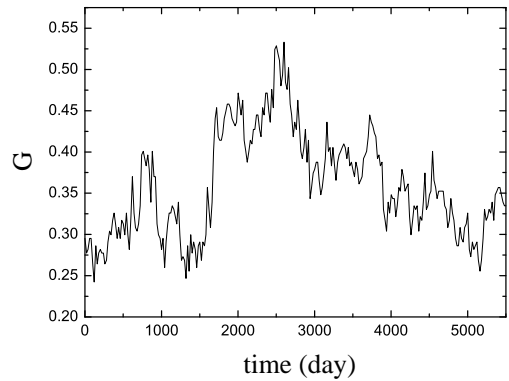} \caption[0]{Plot of the
grouping coefficient for all categories as a function of time.}
\label{grouptime}
\end{figure}

\begin{figure*}[t]
\includegraphics[angle=0,width=\textwidth]{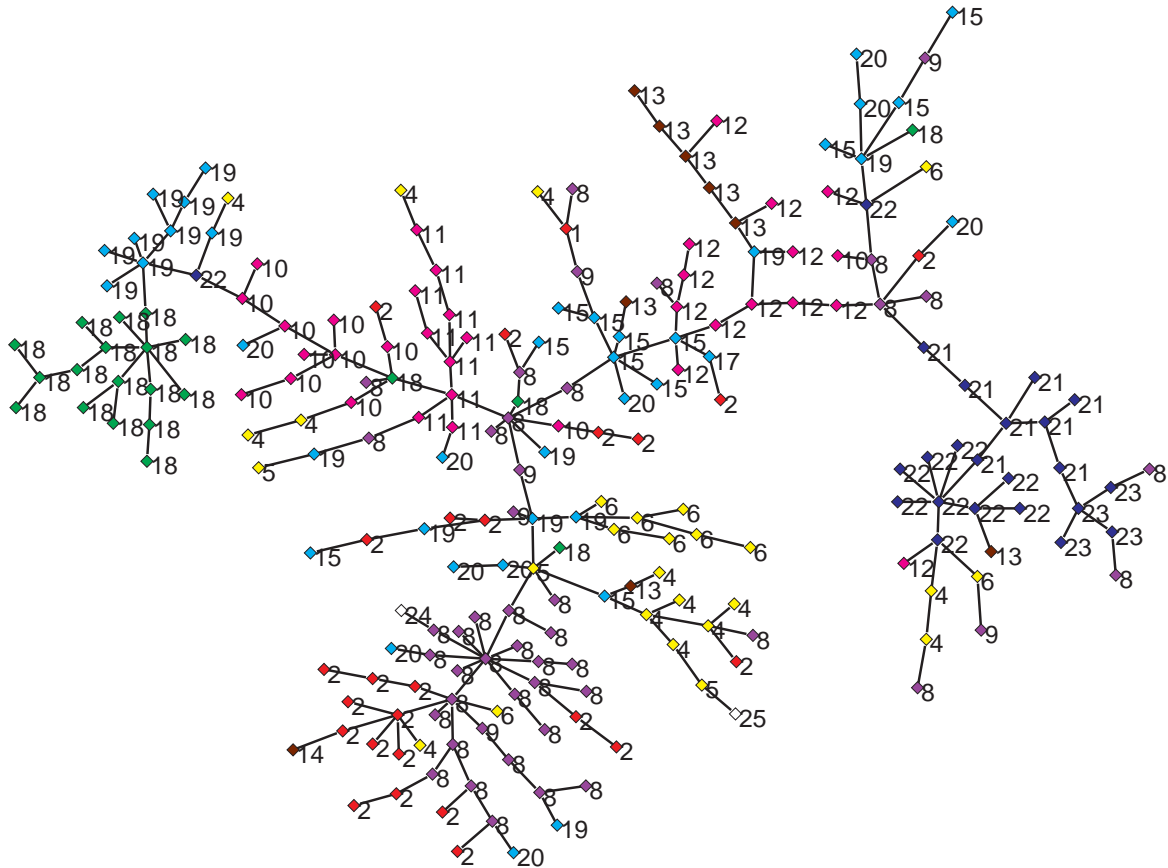}
\caption[0]{The asset tree of the time window from day 2501. In the
US market, the hub is located at the center of the star structure
and clusters are formed through the branches of this star structure
\cite{onnela}. However, we cannot find this property, a comparable
hub and star structure, in the Korean stock market, even though the
grouping coefficient of this period has higher value than the other
periods. The numbers refer to the category number in Table
\ref{category}.} \label{mst}
\end{figure*}

We previously investigated characteristics of the Korean stock
market \cite{wsjung} using a data set from 2001 to 2004, and found
that the market forms clusters when the Morgan Stanley Capital
International (MSCI) index is exploited. However, Fig. \ref{group}
shows that some industry categories can be applied to form the
clusters. We consider the history of the Korean market, including
the globalization effect. Fig. \ref{grouptime} shows the grouping
coefficient $G$ for the whole market as a function of time. This
coefficient is calculated with all of the nodes, and the ratio of
connections between companies in the same category to the total
number of links. Before the mid-1980s, the Korean market had
developed according to a planned economy and had many restrictions
on trading of stocks. At that time, the market was unstable because
of the poor liquidity. This is one possible explanation for lower
value in the early 1980s in Fig. \ref{grouptime}. As the market
prospered in the mid-1980s, clusters of industry categories also
extensively formed.

The 1988 Seoul Olympics Games and the 1997 Asian financial crisis
hastened globalization of the Korean market. In particular,
globalization of the Korean market progressed to synchronization
with external markets. This explains the decreasing coefficient in
Fig. \ref{grouptime} after 1988. The index continues to show a
decreasing trend, which means that the formation of clusters in the
Korean market is related to the MSCI index. The MSCI Korea index has
been calculated from 1988, when the grouping coefficient in Fig.
\ref{grouptime} has almost a maximum value. The MSCI Korea index is
a factor of the Korean market's globalization and market
integration. Because foreign traders strongly influence the Korean
market, the MSCI index is a good reference for their trading
\cite{wsjung}.

\section{Conclusions}
We have studied the Korean stock market network with the daily
closure stock price. The analysis shows that the grouping
coefficient changes with elapsing time. With globalization, the
market is synchronized to external markets, and the number of
clusters of industry categories decreases. Finally, the market forms
clusters according to the MSCI index. We think the tendency of
synchronization will be stronger and the clusters of the MSCI index
or foreign factors will be firmer. Our future research will
determine other properties of the globalization effect with other
statistical analysis in the Korean market.

\begin{acknowledgments}
We thank G. Oh, W. C. Jun, and S. Kim for useful discussions and
support.
\end{acknowledgments}

\end{document}